\documentclass[aps,prd,twocolumn,groupedaddress,showkeys]{revtex4}
\usepackage{graphicx}
\usepackage{dcolumn}
\usepackage{bm}
\usepackage{hyperref} 
\begin{document}

\title{Semiclassical solution of black hole information paradox}

\author{Hrvoje Nikoli\'c}
\affiliation{Theoretical Physics Division, Rudjer Bo\v{s}kovi\'{c} Institute,
P.O.B. 180, HR-10002 Zagreb, Croatia.}
\email{hnikolic@irb.hr}

\date{\today}

\begin{abstract}
We resolve black hole information paradox within semiclassical gravity, in a manner that does not depend on 
details of unknown quantum gravity. Our crucial insight is that outgoing Hawking particles are physical only 
far from the black hole horizon, so they are created far from the horizon and entangled 
with degrees of freedom closer to the horizon. The latter degrees of freedom can be understood as 
quasi-classical coherent states, implying that Hawking radiation is accompanied with additional 
radiation similar to classical radiation by which the black hole loses hair during the classical gravitational 
collapse. The two kinds of radiation are entangled, which resolves black hole information paradox. 
\end{abstract}



\maketitle

\section{Introduction}

\subsection{Motivation}

Black hole information paradox 
\cite{gid,math1,math2,hoss,dundar,harlow,polchinski,chakra,marolf,fabbri}
is one of the greatest unsolved problems in theoretical physics. 
The problem appears within semiclassical theory of gravity 
\cite{bd,mukhanov}, which is an incomplete theory 
where only matter is quantized, while gravity is treated classically.
It is widely believed that the correct solution of black hole information paradox
depends on details of quantum gravity, while our current understanding of 
quantum gravity is still very incomplete, thus making the black hole information paradox 
very difficult to solve. In this paper, however, we argue that the paradox can be solved 
within semiclassical gravity itself, and that details of quantum gravity are not necessary
to understand it. We do not claim that quantum gravity is entirely irrelevant, 
but we argue that its details are not very important.  

\subsection{Where are the Hawking particles created?}
\label{SECwhere}

One of the central questions that we deal with is where are the outgoing Hawking particles created
\cite{unruh_where1,unruh_where2,giddings_where,hod_where}?
The usual picture is that they are created near the horizon. By contrast, we argue that they are created 
far from the horizon, at a distance much larger than the Schwarzschild radius $2M$ 
(with $M$ being the mass of the black hole, and we work in units $\hbar=c=G_N=k_B=1$).

Our argument for this claim is based on considerations of foundations of quantum mechanics itself
\cite{laloe}. The contextuality theorems of quantum mechanics \cite{KS,bell,peres,laloe}
show that it is not consistent to assume that measurements just reveal the values of observables 
that the quantum system possessed before measurements. Instead, it is the measurement that creates
physical reality as we perceive it. Of course, the measurement does not need to be performed by a human made apparatus.
Instead, the measurement of a quantum subsystem is associated with decoherence \cite{decoh1,decoh2}
caused by its environment, which makes the 
density matrix of a subsystem diagonal in a certain ``preferred'' basis determined by interaction with the environment. 

Applying this logic to Hawking radiation, 
it follows that it is inconsistent to speak of outgoing Hawking particles before they are measured
by the environment. The outgoing Hawking particles are defined as positive frequency modes with respect to the 
Schwarzschild time coordinate \cite{bd}, which is a physical time for observers static with respect to the black hole.
Such static observers are approximately inertial when they are far from the horizon, namely at $r\gg 2M$. But at smaller 
distances such observers are non-inertial, so the corresponding ``particles'' can only be physical 
if the environment that measures the particles is itself static and thus non-inertial. In principle it is possible 
to have such a measuring apparatus, but then it measures the particles by the Unruh effect \cite{unruh},
which requires its own source of energy
(such as the fuel of the rocket that keeps a detector hovering at a static position
with respect to the black hole), so such a physical realization of Hawking particles at smaller distances
is compensated by the loss of energy of this source \cite{mochizuki}, which  
does not need to decrease the black hole mass. 
Thus only particles at $r\gg 2M$ are created by measurement without any additional source of energy for the detector,
which makes particles at $r\gg 2M$ much more typical than those at smaller $r$.
Hence, for practical purposes, the creation of Hawking particles is naturally interpreted as something that typically happens 
far from the horizon, at $r\gg 2M$.

\subsection{What are the Hawking particles entangled with?}

With this insight, which is missing in most of the previous approaches to black hole information paradox, 
it is not difficult to understand 
how the black hole information paradox resolves. In the usual picture, 
according to which the outgoing Hawking particles are created near the horizon, they need to be entangled with 
black hole interior degrees of freedom, which leads to the information paradox because the information encoded in the 
interior degrees of freedom cannot escape from the black hole, but also cannot be stored in its interior because the black hole, 
which is shrinking due to the mass loss, cannot keep an arbitrary amount of information.   
But in our picture, where the particles are created far from the horizon, there is no such problem because 
the particles can be entangled with all degrees of freedom at $r$ smaller than that at which the particles are created.
This not only creates much more room for information storage, but also allows information to escape to infinity.
Indeed, we shall argue that Hawking particles are entangled with additional gravitational waves created at those smaller 
distances $r$ outside the horizon, so that both can escape to infinity, thus solving the black hole information paradox.   

\subsection{Organization of the paper}

The rest of the paper is organized as follows. To build intuition for our approach to black hole information paradox,
in Sec.~\ref{SECanal} we consider two simple analogies, 
the hydrogen atom information ``paradox'', and the case of two entangled harmonic oscillators, in settings that resemble 
our view of Hawking radiation. Then in Sec.~\ref{SECsemicl}
we consider the quantum state of matter in a classical black hole background and rewrite
it in terms of physical Hawking particle states at large distances from the horizon, entangled with degrees of freedom at smaller 
distances. In Sec.~\ref{SECzone} we rewrite the state of the latter degrees of freedom in the basis of quasi-classical coherent 
states, 
and argue that radiation carried by Hawking particles is accompanied with an additional quasi-classical radiation,
the latter being similar to radiation produced during the classical gravitational collapse by which a classical black hole
loses classical hair. In Sec.~\ref{SECdisc} we discuss the role of more exotic quantum gravitational objects, 
proposed in the literature as possible solutions of black hole information paradox.      

\section{Simple analogies}
\label{SECanal}

\subsection{Hydrogen atom information paradox}

Consider a slow electron brought close to a proton. Soon the electron will
settle down into the ground state $|\psi_0\rangle$ of the hydrogen atom, independently of the initial electron state.
This already looks like an apparent loss of information about the initial state, but, of course, the information is not lost,
because it is encoded in electromagnetic radiation that electron emits during the process. This is analogous 
to the creation of a stationary black hole during the gravitational collapse, where, 
according to the no hair theorem \cite{hawking-ellis,frolov-novikov}, most information about the initial state is radiated away 
in the form of gravitational waves, leaving a stationary black hole characterized by only a few parameters:
mass, charge, and angular momentum. The hydrogen atom in the ground state is analogous to the stationary black hole, 
while information about the initial state of the electron radiated away by electromagnetic radiation
is analogous to the initial black hole ``hair'' radiated away by gravitational waves. 

Introducing the unit operator in the position basis $1=\int d^3x\,|{\bf x}\rangle\langle{\bf x}|$, 
the ground state can be written as 
\begin{equation}
 |\psi_0\rangle = \int d^3x\,|{\bf x}\rangle\langle{\bf x}|\psi_0\rangle 
 = \int d^3x\,\psi_0({\bf x}) |{\bf x}\rangle .
\end{equation}
Now suppose that the electron, called particle-$A$ for convenience,
is ``measured'' by another particle, called particle-$B$. We assume that particle-$B$ ``measures'' the position of particle-$A$,
so that the full state of the two particles takes the entangled form
\begin{equation}\label{meas}
 |\Psi\rangle \propto \int d^3x\,\psi_0({\bf x}) |{\bf x}\rangle_A |{\bf x}\rangle_B ,
\end{equation}
describing a perfect correlation between positions of the two particles.
Particle-$B$ is in the mixed state
\begin{equation}\label{rhomix}
 \rho_B=\int d^3x\,|\psi_0({\bf x})|^2 [|{\bf x}\rangle \langle{\bf x}|]_B .
\end{equation}

Next suppose that the interaction between the two particles is turned off. The two particles remain entangled.
However, the electron (particle-$A$) is then captured by the proton again, so it ends up in the ground state
$|\psi_0\rangle$. The ground state is unique, so its entropy is zero. Particle-$B$, on the other hand, remains in the 
mixed state with non-zero entropy and it cannot longer be correlated with particle-$A$, because the latter has zero entropy. 
This establishes the hydrogen atom information ``paradox''. 

The solution of the ``paradox'', of course, is that the electron radiates {\em again}
(see Appendix \ref{SECapp_meas}), by the same mechanism as the 
first time.     
This secondary radiation is entangled with particle-$B$, which resolves the ``paradox''.
We shall argue that the black hole information paradox resolves in an analogous way, namely that, in addition to 
the outgoing Hawking particles (analogous to particle-$B$) there is also a secondary radiation of gravitational waves 
(analogous to secondary electromagnetic radiation), so that the outgoing Hawking particles are correlated with 
the secondary radiation.

\subsection{Two harmonic oscillators}

Consider two uncoupled quantum harmonic oscillators, each having the same 
characteristic frequency $\omega$ and an $n$-basis consisting of states of the usual form
\begin{equation}
 |n\rangle=\frac{(a^{\dagger})^n|0\rangle}{\sqrt{n!}} .
\end{equation}
Suppose that the two oscillators are in the entangled state of the thermal form
\begin{equation}\label{psi2ho}
 |\Psi\rangle = {\cal N}\sum_{n=0}^{\infty} e^{-\beta\omega n/2} |n\rangle_A |n\rangle_B ,
\end{equation}
where $\beta$ is the inverse temperature and 
\begin{equation}
{\cal N}=\sqrt{1-e^{-\beta\omega}}
\end{equation}
is the normalization factor. Suppose that the two oscillators are measured in different bases. Oscillator-$B$
is measured in the $n$-basis, while oscillator-$A$ is measured in the basis of canonical coherent states \cite{ballentine} 
$|x,p\rangle\equiv|z\rangle$ obeying $1=\int dz |z\rangle\langle z|$, where $dz\equiv dx\,dp/\pi$.
The basis of coherent states is over-complete and non-orthogonal. Each $|z\rangle$ is a Gaussian in the position and 
momentum space, specified by the average position $x$ and average momentum $p$.
Physically, the measurement in such a basis can be performed by a simultaneous measurement of position and momentum 
\cite{arthurs,stenholm},
resulting in a ``collapse'' into one of the quasi-classical states $|z\rangle$ with uncertain value of both position and momentum, 
such that the product of their uncertainties is minimal, $\Delta x \Delta p= 1/2$ (recall that we work in units $\hbar=1$).
Thus it is natural to write (\ref{psi2ho}) as 
\begin{equation}\label{psi2ho2}
 |\Psi\rangle = \int dz \sum_{n=0}^{\infty} c_n(z) |z\rangle_A |n\rangle_B ,
\end{equation}
where $c_n(z)={\cal N} e^{-\beta\omega n/2} {}_A\langle z|n\rangle_A$. The $|c_n(z)|^2$ is the probability density that 
oscillator-$B$ will be found in the quantum state $|n\rangle_B$, and oscillator-$A$ in the quasi-classical
coherent state $|z\rangle_A$.

\section{Semiclassical black hole}
\label{SECsemicl}

Consider a black hole of mass $M$, described either by Schwarzschild $S$-coordinates
with coordinate singularity at the horizon at $r=2M$
(where $r$ is the usual Schwarzschild radial coordinate),
or by Kruskal $K$-coordinates without a coordinate singularity.
The initial state of the quantum field $\phi(x)$ in the 
classical black hole background is the Kruskal vacuum $|O_K\rangle$ \cite{bd} (known also as Hartle-Hawking vacuum),
naturally associated with $K$-coordinates. It is related to the $S$-vacuum $|O_S\rangle$, naturally associated 
with $S$-coordinates, through the formula \cite{bd}
\begin{equation}\label{0_K}
 |O_K\rangle = \prod_k |\psi_k\rangle ,
\end{equation}
where $|\psi_k\rangle$ is of the form (\ref{psi2ho})
\begin{equation}\label{psik}
 |\psi_k\rangle = {\cal N}_k\sum_{n_k=0}^{\infty} e^{-\beta\omega_k n_k/2} |n_k,n_k\rangle ,
\end{equation}
\begin{equation}
 |n_k,n_k\rangle = \frac{ (b_{1k}^\dagger)^{n_k}  (b_{2k}^\dagger)^{n_k}  |O_S\rangle}{n_k!} .
\end{equation}
Here $\beta=8\pi M$ is the inverse black hole temperature, $b_{1k}^\dagger$ creates ``particles''
in zone-1 (black hole interior $r\leq 2M$), and $b_{2k}^\dagger$ creates ``particles''
in zone-2 (black hole exterior $r> 2M$). 

Before proceeding, a few notes are in order considering the choices of vacua above. 
$K$-coordinates and $S$-coordinates are both analogous to Minkowski coordinates
in flat spacetime, but in a different sense. $K$-coordinates are analogous to Minkowski 
coordinates in the sense that they do not contain a coordinate singularity at the horizon.
In this sense $S$-coordinates are analogous to Rindler coordinates in flat spacetime, 
because both contain a coordinate singularity at the horizon.  
But $S$-coordinates are analogous to Minkowski coordinates in the sense that 
the components of the metric tensor in Minkowski coordinates look like that 
in $S$-coordinates at $r\gg 2M$.
Hence the assumption that the initial state is the $K$-vacuum corresponds
to the assumption that there was no physical Minkowski particles before 
the black hole was formed. But after the formation of the black hole,
the physical particles at $r\gg 2M$ are to be defined with respect to the $S$-vacuum.
Of course, this refers to the Schwarzschild black hole, which does not have charge 
and angular momentum. For more general black holes, with charge and angular momentum,
one would need to define analogous vacua with respect to coordinates that generalize 
$K$-coordinates (by not having coordinate singularities) and $S$-coordinates (by 
having components of the metric tensor far from the black hole like that in Minkowski coordinates),
but in this paper only the simplest Schwarzschild black hole will be considered explicitly. 

At $r\gg 2M$, the ``particles'' created by $b_{2k}^\dagger$ behave as ordinary particles in quantum field theory in 
flat Minkowski spacetime.
For that reason, those ``particles'' are considered in the literature to be the actual physical particles, 
so (\ref{psik}) is interpreted
as physical particle creation in zone-2. However, the black hole interior in zone-1 has a limited number of degrees of freedom,
i.e. limited entropy, and it turns out that it does not have enough entropy to explain the entropy of particles created in zone-2,
through the entanglement between zone-2 and zone-1. This is the essence of black hole information paradox 
\cite{gid,math1,math2,hoss,dundar,harlow,polchinski,chakra,marolf,fabbri}.

Our crucial new insight is the following. As we already explained in Sec.~\ref{SECwhere},
the particles created by $b_{2k}^\dagger$ are physical at very large distances $r\gg 2M$, 
where the spacetime is essentially Minkowski spacetime, but they are not physical at intermediate distances $r$.
Hence we split zone-2 into two zones, called zone-$A$ and zone-$B$, defined as follows:
\begin{equation}
\begin{array}{l}
 \text{zone-}A: \; 2M<r\le R', 
\\
 \text{zone-}B: \; r>R', 
\end{array}
\end{equation}
where $R'$ is some fixed large radius $R'\gg 2M$. 
(The precise value of $R'$ is not important. The two zones with a sharp boundary serve as a simplified model, 
while in a more realistic model there would be no sharp boundary between the two zones.)
Hence, as shown by (\ref{bogol1}) in Appendix \ref{SECbogol}, 
the creation operator $b_{2k}^\dagger$ can be decomposed into two operators 
\begin{equation}\label{b2k}
 b_{2k}^\dagger=\alpha_{Ak}^*b_{Ak}^\dagger + \alpha_{Bk}^*b_{Bk}^\dagger ,
\end{equation}
which is a ``trivial'' Bogoliubov transformation, in the sense that it does not mix creation and destruction operators.
Since the new creation and destruction operators must satisfy the usual commutation relations 
$[b_{Ak}, b_{Ak}^\dagger]=1$, $[b_{Bk}, b_{Bk}^\dagger]=1$, etc., the Bogoliubov coefficients must satisfy
\begin{equation}\label{sum_alpha=1}
 |\alpha_{Ak}|^2+|\alpha_{Bk}|^2 =1. 
\end{equation}
The binomial theorem applied to (\ref{b2k}) implies
\begin{eqnarray}
 &  (b_{2k}^\dagger)^{n_k}  = (\alpha_{Ak}^*b_{Ak}^\dagger + \alpha_{Bk}^*b_{Bk}^\dagger)^{n_k} =&
\\
& \displaystyle\sum_{l_k=0}^{n_k} \left( \begin{array}{c} n_k \\ l_k \end{array} \right) 
(\alpha_{Ak}^*)^{n_k-l_k} (\alpha_{Bk}^*)^{l_k}  (b_{Ak}^\dagger)^{n_k-l_k}  (b_{Bk}^\dagger)^{l_k} ,
\nonumber
\end{eqnarray}
where 
\begin{equation}
 \left( \begin{array}{c} n_k \\ l_k \end{array} \right) =\frac{n_k!}{(n_k-l_k)!l_k!}
\end{equation}
are the binomial coefficients. Hence, using also 
\begin{eqnarray}
& (b_{1k}^\dagger)^{n_k} (b_{Ak}^\dagger)^{n_k-l_k} (b_{Bk}^\dagger)^{l_k} |0_S\rangle = &
\nonumber \\
&\sqrt{n_k! (n_k-l_k)! l_k!}\; |n_k\rangle_1 |n_k-l_k\rangle_A |l_k\rangle_B , &
\end{eqnarray}
the state (\ref{psik}) can finally be written as 
\begin{equation}\label{psik2}
 |\psi_k\rangle = \sum_{n_k=0}^{\infty} \sum_{l_k=0}^{n_k} C_{n_kl_k} |\psi_{n_kl_k}\rangle_{1\cup A} |l_k\rangle_B ,
\end{equation}
where
\begin{equation}\label{Cnklk}
C_{n_kl_k} = {\cal N}_k e^{-\beta\omega_k n_k/2} \sqrt{\left( \begin{array}{c} n_k \\ l_k \end{array} \right)}
(\alpha_{Ak}^*)^{n_k-l_k} (\alpha_{Bk}^*)^{l_k} ,
\end{equation}
\begin{equation}
 |\psi_{n_kl_k}\rangle_{1\cup A} = |n_k\rangle_1 |n_k-l_k\rangle_A .
\end{equation}

The upshot of this calculation is Eq.~(\ref{psik2}), which shows that physical particles in zone-$B$ are entangled 
with the states in zone-$(1\!\cup\! A)$, the union of zone-1 and zone-$A$. Since $R'\gg 2M$, 
zone-$(1\!\cup\! A)$ is much larger than the black hole interior zone-1. 
Moreover, most of the zone-$(1\!\cup\! A)$ is outside of the black hole.
Thus the zone-$(1\!\cup\! A)$ contains a plenty of space for storing entropy, and most of it is not hidden behind the horizon. 
Since the ``particles'' in zone-$(1\!\cup\! A)$ are not physical particles, the physical particle creation 
in our model can be considered to occur at the sphere of radius $R'\gg 2M$ 
\footnote{This picture, of course, should be taken with a grain of salt, because in a more realistic model there would be no sharp 
boundary between the two zones.}.
The entanglement entropy of particles created at $R'$ 
can easily be explained through entanglement with degrees of freedom in zone-$(1\!\cup\! A)$, 
which, in principle, resolves the black hole information paradox.

Furthermore, since zone-$B$ is naturally defined as extending to $r\to \infty$, its volume $V_B$ is essentially infinite.
Hence (\ref{aVAB}) implies $|\alpha_{Ak}|^2/|\alpha_{Bk}|^2=0$, which, in combination with (\ref{sum_alpha=1}), 
implies $|\alpha_{Bk}|^2=1$, $|\alpha_{Ak}|^2=0$. Hence $\alpha_{Bk}^*=e^{-i\varphi_{Bk}}$, $\alpha_{Ak}^*=0$, 
and only $n_k-l_k=0$ contributes to (\ref{Cnklk}). Thus, absorbing the phase $e^{-i\varphi_{Bk}}$ into a redefinition of
$|l_k\rangle_B$, Eq.~(\ref{psik2}) simplifies to
\begin{equation}\label{psik3}
 |\psi_k\rangle = {\cal N}_k \sum_{n_k=0}^{\infty} e^{-\beta\omega_k n_k/2} |\psi_{n_kn_k}\rangle_{1\cup A} |n_k\rangle_B .
\end{equation}
This looks very similar to (\ref{psik}), which, of course, is not a coincidence because 
(\ref{psik}) are (\ref{psik3}) are two representations of the same state.
The difference is that only (\ref{psik3}) is expressed in terms of physical particles in zone-$B$, while  
(\ref{psik}) is expressed in terms of particles in zone-2, which are not entirely physical.
Again, (\ref{psik3}) shows that the physical particles in zone-$B$ are entangled with degrees of freedom
in the entire zone-$(1\!\cup\! A)$, which is sufficiently large to resolve the black hole information paradox.

Finally note that, even though $R'\gg 2M$, it would not make much sense to take the limit $R'\to\infty$. 
In this limit zone-$A$ would take all of the black hole exterior, so zone-$B$ would not exist.
This would mean that physical particles could not exist at any finite distance from the black hole.
And this would refer not only to Hawking particles, but to all particles that we usually describe by 
quantum field theory in flat spacetime, such as those that we observe by standard detectors in particle physics, e.g., at CERN. Infinite $R'$ would contradict the fact that we, of course, do observe such particles, despite the existence of black holes at finite distance from Earth. 
Hence we must render $R'$ finite.  

\section{What happens in zone-$(1\!\cup\! A)$?}
\label{SECzone}

So far we explained that zone-$(1\!\cup\! A)$, 
bounded by the surface of large radius $R'\gg 2M$, is sufficiently large to accommodate all the entropy needed to resolve
the black hole information paradox. Nevertheless, since the particle states in that zone are not physical particles, 
the question is how to understand that zone in terms of physical objects?
In other words, what an inertial observer in that zone would observe?

In principle, this question should be answered by the theory of decoherence \cite{decoh1,decoh2}. 
In quantum field theory, depending on details of interaction with the environment, decoherence 
explains why the quantum states are typically observed either as states with definite number of particles, 
or as quasi-classical coherent states resembling classical fields \cite{zeh,zurek}.
In particular, decoherence explains under which conditions acceleration of an environment creates particles    
\cite{hu,audretsch,kok}, rather than quasi-classical coherent states of fields. 
Since, as we have explained, there is typically no creation of physical particles in zone-$(1\!\cup\! A)$,
it is natural to assume that quantum states manifest themselves as quasi-classical coherent states of fields.   
This refers to matter fields, but also to gravitational fields. Hence, in principle, some elements of quantum 
gravity also need to be taken into account, but here we do it in a minimal manner that does not depend on 
details of the quantum theory of gravity. We assume that the quantum state of matter and gravity has a form 
similar to (\ref{0_K}), 
\begin{equation}\label{M}
|M\rangle = \prod_k |M,k\rangle , 
\end{equation}
where $M$ denotes that the state describes a black hole of mass $M$,
and the state $|M,k\rangle$ is a quantum-gravitational extension of (\ref{psik2})
\begin{equation}\label{psik4}
 |M,k\rangle = \sum_{n_k=0}^{\infty} \sum_{l_k=0}^{n_k} C_{n_kl_k} |M,n_k,l_k\rangle_{1\cup A} |l_k\rangle_B .
\end{equation}
Then we assume that, in zone-$(1\!\cup\! A)$,
there is an over-complete basis of quasi-classical coherent states $|Z_{1\cup A}\rangle$ 
for both gravitational and matter fields, 
$|Z_{1\cup A}\rangle=|Z_{\rm gravitational},Z_{\rm matter}\rangle$, obeying 
\begin{equation}\label{cohZ}
1_{1\cup A}=\int [dZ_{1\cup A}]\, |Z_{1\cup A}\rangle \langle Z_{1\cup A}| .
\end{equation}
Hence (\ref{M}) can be written as 
\begin{equation}\label{M2}
 |M\rangle = \int [dZ_{1\cup A}] \prod_k \sum_{n_k=0}^{\infty} \sum_{l_k=0}^{n_k} 
\Psi_{n_kl_k}[Z_{1\cup A}] \;  |Z_{1\cup A}\rangle |l_k\rangle_B ,
\end{equation}
where
\begin{equation}
\Psi_{n_kl_k}[Z_{1\cup A}]=C_{n_kl_k} \langle Z_{1\cup A}|M,n_k,l_k\rangle_{1\cup A} .
\end{equation}
If we restrict it to the terms $l_k=n_k$ as in (\ref{psik3}), then (\ref{M2}) simplifies to
\begin{equation}\label{M3}
 |M\rangle = \int [dZ_{1\cup A}] \prod_k \sum_{n_k=0}^{\infty}  
\Psi_{n_kn_k}[Z_{1\cup A}] \;  |Z_{1\cup A}\rangle |n_k\rangle_B .
\end{equation}
The $\prod_k|\Psi_{n_kn_k}[Z_{1\cup A}]|^2$ is the probability density that there will be $n_k$ particles (in the $k$-modes)
in zone-$B$, and that the fields will be in the quasi-classical coherent state $|Z_{1\cup A}\rangle$
in zone-$(1\!\cup\!A)$.  

Without a full theory of quantum gravity, we cannot specify the coherent states $|Z_{1\cup A}\rangle$ explicitly.
Nevertheless, their essential properties can easily be inferred by heuristic arguments. It is natural to expect that 
a typical coherent state resembles a classical configuration of gravitational and matter fields. 
Hence, a typical coherent state behaves approximately classically. 
In other words, it behaves as a classical hair in zone-$(1\!\cup\!A)$. But we know from  
classical no hair theorems \cite{hawking-ellis,frolov-novikov} that classical hair soon gets radiated away, in terms 
of gravitational (and matter) waves. Thus our approach predicts that there are {\em two} kinds of radiation
from the black hole. First there are Hawking particles created in zone-$B$, and second there is a
quasi-classical radiation of gravitational and matter waves created in zone-$A$. 
The two kinds of radiation are entangled with each other, which resolves the black hole information paradox.

\section{Discussion and conclusion}
\label{SECdisc}

\subsection{Quantum gravity exotics?}

Our approach, which does not depend on details of quantum gravity, 
predicts that Hawking radiation is entangled with additional quasi-classical radiation, the latter being similar
to classical radiation produced by classical gravitational collapse. In this sense, our approach does not predict 
any exotic quantum gravitational objects, such as firewalls \cite{AMPS}, fuzzballs \cite{mathur}, wormholes \cite{er=epr},
islands \cite{island}, baby-universes \cite{gid}, white holes \cite{rovelli_tunnel}, supertranslation hair \cite{softhair}
and gravitational crystals \cite{nik_cryst},
that have been proposed in other approaches to the black hole information paradox. 
But our approach does not exclude such exotics either. Instead, all such exotic states can be implicitly included 
in the set of all coherent states contributing to (\ref{cohZ}). In principle, there may be a non-zero probability 
for formation of various kinds of such exotic states, depending on details of the unknown quantum theory of gravity.
Perhaps even all exotic objects mentioned above have a non-zero probability. But if more than one kind of 
states $|Z_{1\cup A}\rangle$ contributes, the relevant question is which of these contributions dominates. 
Such a question cannot be definitely answered without a full theory of quantum gravity. Nevertheless,
the most conservative, and perhaps the most plausible, is the scenario in which (\ref{M3}) is dominated 
by the configurations that maximally resemble the known field configurations in classical general relativity, 
namely classical hair radiated away by classical mechanisms, as we explained in Sec.~\ref{SECzone}.

\subsection{Tunnelling picture}

The description of Hawking radiation in this paper is based on Bogoliubov transformation, 
which describes the radiation as a state in the Hilbert space and makes 
the quantum entanglement explicit. However, such a description lacks a simple physical intuitive picture 
of particle creation. Alternatively, to make particle creation more intuitive, Hawking radiation
can be described as quantum tunnelling \cite{pw,zhang}. Even though quantum tunnelling in this context 
is not 
explicitly formulated in the Hilbert space, so the entanglement is not seen explicitly, 
a possible resolution of the 
information paradox can be formulated in terms of energy correlations between particles that tunnel at different 
times \cite{zhang}.

In our resolution of the paradox there is no correlation between particles created 
at different times, but there is a correlation between particles and quasi-classical radiation 
created at about the same time. In principle, this could also be understood in terms of tunnelling, 
in a process in which both the radiation and the particles tunnel. Since the quasi-classical radiation is created 
in zone-$A$, in the tunnelling picture it tunnels through the horizon. By contrast, since the particle 
is created in zone-$B$, the whole zone-$A$ can be thought of as a classically forbidden zone for the 
particle, so the particle tunnels not only through the horizon but also through zone-$A$.
An explicit technical analysis of such a tunnelling picture is beyond the scope of the present paper,
but it can be an interesting challenge for the future work. 

\subsection{Conclusion}

The crucial ingredient of our approach is the idea that Hawking particles are physical 
only at large distances $r\gg 2M$ from the black hole, where inertial observers are approximately static with respect 
to the black hole. At smaller distances it is more physical to describe physics in terms of different objects, 
that can be entangled with Hawking particles at large distances. Those different objects are most naturally described in terms 
of quasi-classical fields, leading to the picture in which black hole produces two kinds of radiation,
Hawking particles and quasi-classical radiation similar to the radiation produced during the classical black hole collapse.
The entanglement between the two kinds of radiation resolves black hole information paradox. 

Such resolution of the paradox does not depend on any details of quantum gravity and does not involve 
any exotic phenomena that would contradict general expectations from semiclassical gravity.
In particular, the resolution does not involve exotic objects near the horizon,
such as firewalls and fuzzballs, so 
a freely falling observer near the horizon will not observe anything out of the ordinary.
Moreover, while in the standard approach, where particles are created near the horizon, 
one can wonder whether the particle creation will be observed by a freely falling observer,
in our approach it is clear that a freely falling observer near the horizon will not observe any Hawking particles, because in our approach Hawking particles are unphysical in the whole 
zone-$(1\!\cup\! A)$, including the region near the horizon.

To conclude, we believe that our resolution of black hole information paradox 
offers a very plausible picture worth of further research.
        

\appendix

\section{The effect of measurement on the quantum state}
\label{SECapp_meas}

In this Appendix we present the basic scheme describing how measurement affects the quantum state. 
For illustration purposes we describe it
for the case of position measurement of electron in the hydrogen atom, but essentially the same scheme works 
also for measurements in semiclassical and quantum gravity.

To give an operational meaning to the mixed state (\ref{rhomix}), the position of particle-$B$ must actually be measured
by a macroscopic measuring apparatus M. Hence the state (\ref{meas}) generalizes to 
\begin{equation}\label{meas2}
 |\Psi\rangle \propto \int d^3x\,\psi_0({\bf x}) |{\bf x}\rangle_A |{\bf x}\rangle_B  |{\rm M}_{\bf x}\rangle,
\end{equation}
where $|{\rm M}_{\bf x}\rangle$ is the state of the measuring apparatus M corresponding to the measurement outcome ${\bf x}$.
So, if the measurement outcome of a single measurement turns out to be ${\bf x}$, 
we have an effective measurement-induced ``collapse'' 
\begin{equation}\label{meas3}
 |\Psi\rangle \to |{\bf x}\rangle_A |{\bf x}\rangle_B  |{\rm M}_{\bf x}\rangle .
\end{equation}
Hence the electron is in the state $|{\bf x}\rangle_A$, namely, in a state with a well defined position 
different from the ground state $|\psi_0\rangle_A$. The state $|{\bf x}\rangle_A$ is not stationary, so 
the electron soon settles down into the ground state and produces electromagnetic radiation during this process. 
The state of radiation depends
on the pre-radiation state $|{\bf x}\rangle_A$, so radiation is correlated with particle-$B$ in the state $|{\bf x}\rangle_B$.

\section{Bogoliubov coefficients}
\label{SECbogol}

The field operator in zone-2 can be expanded as \cite{bd}
\begin{equation}\label{ap1}
 \phi_2(x)=\sum_k \left( b_{2k} f_{2k}(x) + b_{2k}^\dagger f_{2k}^*(x) \right) .
\end{equation}
Alternatively, it can also be expanded as
\begin{eqnarray}\label{ap2}
 \phi_2(x) &=& \sum_k \left( b_{Ak} f_{Ak}(x) + b_{Ak}^\dagger f_{Ak}^*(x) \right)
\nonumber \\
& & + \sum_k \left( b_{Bk} f_{Bk}(x) + b_{Bk}^\dagger f_{Bk}^*(x) \right),
\end{eqnarray}
where the modes $f_{Ak}(x)$ and $f_{Bk}(x)$ have support in zone-$A$ and zone-$B$, respectively.
In principle one would need to construct the modes $f_{Ak}(x)$ and $f_{Bk}(x)$ explicitly, 
but that would be complicated. Fortunately, a lot can be concluded 
without an explicit construction.

Consider the Klein-Gordon scalar products $(f_{2k},f_{Ak'})$, $(f_{2k},f_{Bk'})$,
$(f_{2k},f_{Ak'}^*)$ and $(f_{2k},f_{Bk'}^*)$, where the Klein-Gordon scalar product is defined as \cite{bd}
\begin{equation}\label{KG}
 (f,h)=i\int_{\Sigma} d\Sigma^{\mu}f^*(x)  \!\stackrel{\leftrightarrow\;}{ \partial_\mu }\! h(x) ,
\end{equation}
with $d\Sigma^{\mu}$ being the integration measure over a 3-dimensional spacelike hypersurface $\Sigma$.
The typical wavelengths associated with Hawking radiation are of the order of $2M$.
Since the sizes of zone-$A$ and zone-$B$ are much larger than those typical wavelengths,
for typical $k$ and $k'$ the Klein-Gordon scalar products above are negligible for $k\neq k'$.
Hence only the $k=k'$ terms are non-negligible,  
so $(f_{2k}, \phi_2)$ applied to both (\ref{ap1}) and (\ref{ap2}) gives
\begin{equation}\label{bogol0}
 b_{2k}= \alpha_{Ak}b_{Ak} + \beta_{Ak}b_{Ak}^\dagger + \alpha_{Bk}b_{Bk} + \beta_{Bk}b_{Bk}^\dagger ,
\end{equation}
where
\begin{eqnarray}\label{bogol}
 \alpha_{Ak}=(f_{2k},f_{Ak}), \;\; \beta_{Ak} = (f_{2k},f_{Ak}^*) ,
\nonumber \\
\alpha_{Bk}=(f_{2k},f_{Bk}), \;\; \beta_{Bk} = (f_{2k},f_{Bk}^*) ,
\end{eqnarray}
are the Bogoliubov coefficients.
For the same reason, the scalar products $(f_{2k},f_{Ak}^*)$ and $(f_{2k},f_{Bk}^*)$, i.e., 
the $\beta$-coefficients in (\ref{bogol}), are also negligible, so (\ref{bogol0}) reduces to
\begin{equation}\label{bogol1}
 b_{2k}= \alpha_{Ak}b_{Ak} + \alpha_{Bk}b_{Bk}. 
\end{equation}

Furthermore, the order of magnitudes of the $\alpha$-coefficients can also be estimated.
Since the Klein-Gordon scalar product (\ref{KG}) involves an integration over the 3-volume, and since 
norms of all modes must be unit, $(f_{2k},f_{2k})=(f_{Ak},f_{Ak})=(f_{Bk},f_{Bk})=1$, 
these modes must have a normalization factor proportional to $1/\sqrt{\text{3-volume}}$, namely
\begin{equation}
 f_{2k}\propto \frac{1}{\sqrt{V_2}} , \;\;\; f_{Ak}\propto \frac{1}{\sqrt{V_A}} , \;\;\; f_{Bk}\propto \frac{1}{\sqrt{V_B}} . 
\end{equation}
Hence
\begin{equation}
 |\alpha_{Ak}| \sim \frac{V_A}{\sqrt{V_2V_A}}, \;\;\; |\alpha_{Bk}| \sim \frac{V_B}{\sqrt{V_2V_B}} , 
\end{equation}
where the numerators $V_A$ and $V_B$ arise from integration in (\ref{KG}) over the support of $f_{Ak}(x)$ and $f_{Bk}(x)$, 
respectively. Hence
\begin{equation}\label{aVAB}
 \frac{|\alpha_{Ak}|^2 }{|\alpha_{Bk}|^2} \sim \frac{V_A}{V_B} .
\end{equation}

\end{document}